# Steel-Based Electrocatalysts for Efficient and Durable Oxygen Evolution in Acidic Media

*Helmut Schäfer*[a]*, *Karsten Küpper*[a, b], *Mercedes Schmidt*[a], *Klaus Müller-Buschbaum*[c], *Johannes Stangl*[c] *Diemo Daum*[d], *Martin Steinhart*[a], *Christine Schulz-Kölbel*[a], *Weijia Han*[a], *Joachim Wollschläger*[a, b], *Ulrich Krupp*[e], *Peilong Hou*[a], *and Xiaogang Liu*[f]

[a]*Institute of Chemistry of New Materials and Center of Physics and Chemistry of New Materials, Universität Osnabrück, Barbarastrasse 7, 49076 Osnabrück, Germany*

[b]*Department of Physics, Universität Osnabrück, Barbarastraße 7, 49069 Osnabrück, Germany*

[c]*Institute of Inorganic Chemistry Julius-Maximilians-Universität Würzburg*
*Am Hubland, D-97074 Würzburg, Germany*

[d]*Faculty of Agricultural Science and Landscape Architecture, Laboratory of Plant Nutrition and Chemistry, Osnabrück University of Applied Sciences, Am Krümpel 31, 49090 Osnabrück, Germany*

[e]*Institute of Materials and Structural Integrity University of Applied Sciences Osnabrueck, Albrechtstraße 30, 49076 Osnabrück, Germany*

[f]*Department of Chemistry, Faculty of Science, National University of Singapore, 3 Science Drive 3, Singapore 117543.*

*E-mail: helmut.schaefer@uos.de.

High overpotentials, particularly an issue of common anode materials, hamper the process of water electrolysis for clean energy generation. Thanks to immense research efforts up to date oxygen evolution electrocatalysts based on earth-abundant elements work efficiently and stably in neutral and alkaline regimes. However, non-noble metal-based anode materials that can withstand low pH regimes are considered to be an indispensable prerequisite for the water splitting to succeed in the future.

All oxygen evolving electrodes working durably and actively in acids contain Ir at least as an additive. Due to its scarcity and high acquisition costs noble elements like Pt, Ru and Ir need to be replaced by earth abundant elements. We have evaluated a Ni containing stainless steel for use as an oxygen-forming electrode in diluted $H_2SO_4$. Unmodified Ni42 steel showed a significant weight loss after long term OER polarization experiments. Moreover, a substantial loss of the OER performance of the untreated steel specimen seen in linear sweep voltammetry measurements turned out to be a serious issue. However, upon anodization in LiOH, Ni42 alloy was rendered in OER electrocatalysts that exhibit under optimized synthesis conditions stable overpotentials down to 445 mV for 10 mA $cm^{-2}$ current density at pH 0. Even more important: The resulting material has proven to be robust upon long-term usage (weight loss: 20 $\mu g/mm^2$ after 50 ks of chronopotentiometry at pH 1) towards OER in $H_2SO_4$. Our results suggest that electrochemical oxidation of Ni42 steel in LiOH (sample Ni42Li205) results in the formation of a metal oxide containing outer zone that supports solution route-based oxygen evolution in acidic regime accompanied by a good stability of the catalyst.

## Table of contents

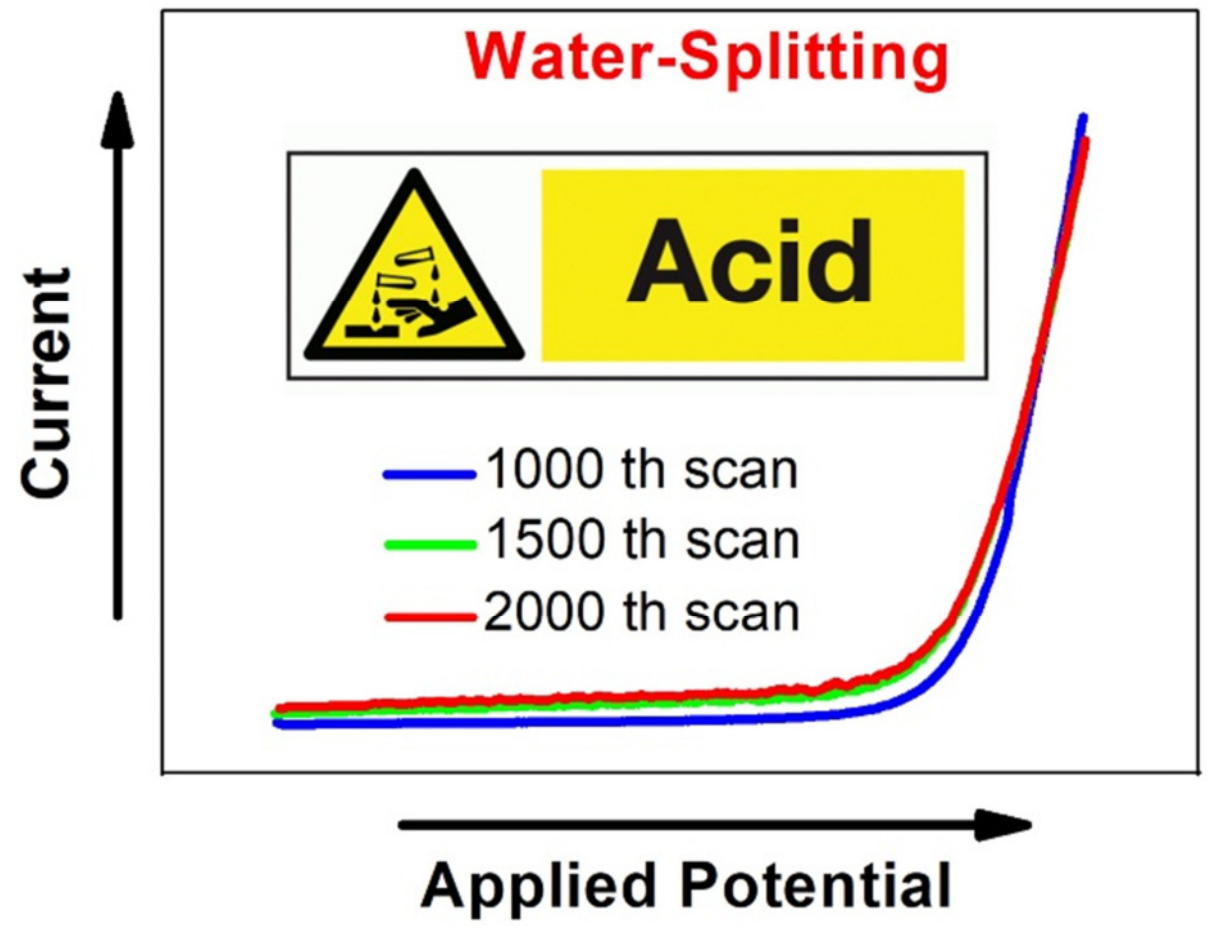

Ni42 alloy was upon electro-oxidation in LiOH rendered in an OER electrocatalysts that efficiently and durably works in acidic regime.

## Introduction

The limited availability of fossil fuels like oil coal and gas forced scientist and engineers to look for new efficient solar to fuel conversion routes and new storage possibilities capable for the storage of solar energy [1-5]. Pumped hydro/turbines are the most efficient electrical storage/production solution, but it is not implementable everywhere and most relevant sites are already saturated. Hydrogen, a clean energy carrier which shows among the known fuels the highest specific energy density is easily accessible via water-splitting by exploration of solar energy [6-15]. However, electrochemically initiated cleavage of water molecules cannot be performed at the theoretically required (thermodynamic) potential as derived from electrochemical series, *i.e.* the efficiency of the electrochemical approach suffers from high overpotentials particularly on the anode side. Efficiency and stability of common electrode materials exploited as anodes in water electrolysis cells critically depend on the pH value of the electrolyte. Both performance and durability of electrocatalysts based on earth- abundant elements toward OER in neutral and alkaline regime has been significantly improved [6, 16-28] and is not any more a crucial point. Alkaline water electrolysis can therefore be seen as the most widespread and mature technology. Since water electrolysis cells received particular attention as a component of a plant located close to offshore wind parks, it must cope with many special requirements as there are: resistance against frequently occurring changes of the current load, insensitiveness to interference and ease of maintenance.

Unfortunately, alkaline electrolyzer technique cannot be set well on greatly fluctuating load whereas Proton exchange membrane (PEM) electrolyzers are seen as the method of choice for the storage of renewable energy characterized by high dynamics [6]. In addition, PEM electrolyzers benefit from higher gas purity and higher efficiency. However, the operation of these electrolyzers requires a low pH condition. It is easy to imagine that the development of anodes solely consisting of earth-abundant elements suitable for oxidative water-splitting of acids is a challenging task, especially so when taking into consideration the strong positive potentials encountered in water electrolyzers [29].

Transition metal oxides such as Ni-oxide, Co-oxide, and Ir-oxide, have been intensively investigated for use as OER electrocatalysts in 1 M $H_2SO_4$ [30]. Recent studies showed that every compound except for IrOx was found to be unstable under oxidative conditions in acidic solutions [31]. Generally, $IrO_x$ and $RuO_x$ are currently the only known materials that can reach 5 mA/cm$^2$ with overpotentials less than 750 mV in acidic electrolyte [31, 32] with sufficient stability. Addition of a “diluting agent”, *e.g.* other metal oxides represents one reasonable and very promising strategy to reduce the price and improve the OER electrocatalytic properties of $IrO_2$-, $RuO_2$- or $RuO_2$-$IrO_2$ based electrocatalysts [33-39]. Because of high acquisition costs and scarcity of both Ir and Ru, future efforts are aimed at the total replacement of these elements by abundant cheap elements. Therefore, the development of electrocatalysts solely based on non-precious metals suitable for robust and efficient anodic water-splitting in acidic regime is of highest interest [31], and certainly represents the shared view of many research groups.

Flat heterogeneous catalysts based on closed packed metal atoms can be seen as industrially relevant, (technical) ones[40]. Traditionally we use surface modified steel as electrode material [20, 29, 41, 42, 43] for electrocatalytically initiated splitting of water due to their high efficiency and stability and because of the fact that they are inexpensive and easily accessible. Moreover, three-dimensional structures based on stainless steel are more or less omnipresent like *e.g.* stainless steel sponges [44], felts [45]-tangles, scrubbers [46] and can be used "as is" for water-splitting [46].

Until recently, we have failed to render steel in an electrocatalyst that can be considered as competitive to Ir-Ru-containing OER electrocatalysts for water-splitting in acids[29, 47]. In this report, we evaluate the suitability of untreated- as well as surface modified-steel as OER electrocatalysts for the water splitting reaction performed at pH 0 and pH 1. Ni42 steel, basically consisting of Fe, Ni and Mn, electro-oxidized in LiOH showed high stability towards long term usage as OER electrocatalyst in 0.05 M $H_2SO_4$ and was found to be highly competitive to recently developed and significantly more expensive OER electrocatalysts that work in acidic environment.

## Experimental

### Sample preparation

**Samples made from untreated steels (Samples Ni42).** Samples with a total geometry of 70x10x1,5 mm were constructed from 1,5 mm thick sheets consisting of AISI Ni42 steel. AISI Ni42 steel was purchased from Schmiedetechnik Faulenbach, Wiehl, Germany. Pre-treatment: The surface of the metal was cleaned intensively with ethanol and polished with grit 240 SiC sanding paper. Afterwards the surface was rinsed intensively with deionized water and dried under air for 100 min at 50 °C.

**Ni42Li127, Ni42Li205, Ni42Li300 sample series.** Samples with a total geometry of 80x10x1,5 mm were constructed from 1,5 mm thick AISI Ni 42 steel. Pre-treatment: Prior to each surface modification, the surface of the metal was cleaned intensively with ethanol and polished with grit 240 SiC sanding paper. Afterwards the surface was rinsed intensively with deionized water and dried under air for 100 min at 50 °C. The weight was determined using a precise balance (Sartorius 1712, 0.01 mg accuracy) prior to electro-activation. A two-electrode set-up consisting of the steel sample as working electrode (WE) and a platinum wire electrode (4x5 cm) used as counter electrode (CE) was exploited for the electro-oxidation. The WE (anode) was immersed exactly 2.1 cm deep (4.5 $cm^2$ geometric area), and the CE (cathode) was completely immersed into the electrolyte.

The electrolyte was prepared as follows: In a 1000 mL glass beaker, 86.21 g (3.6 mol) of LiOH (VWR, Darmstadt, Germany) was added to 600 mL of distilled water under stirring and the resulting mixture was filled up to a total volume of 750 mL with the final concentration of 4.8 M/L LiOH. The solution was allowed to cool down to 23°C before usage. The anodization was performed in a glass beaker (150 mL) in 120 mL of the electrolyte under stirring (450 r/min) using a magnetic stirrer and a stirring bar (21 mm in length, 6 mm in diameter). The distance between WE and CE was adjusted to 6 mm. A power source (Electra Automatic, Vierssen, Germany) EA-PSI 8360-15T which allows to deliver a constant voltage even at strongly changing current loads was used for the electrochemical oxidation. The procedure was carried out in current controlled mode. The current was set to 8 A according to 1778 $mA/cm^2$ current density. The voltage varied during the electro-activation. At the beginning of the experiment it amounted to 6-6.3 V but was reduced within the first 100 min of electro-oxidation to around 4.5 V. The electro-oxidation procedure was stopped after 127 min (Ni42-Li127), 205 min (Ni42Li205), 300 min (Ni42Li300) respectively. The temperature of the electrolyte increased within the first 30 min and reached a value of 323 K which was found to be constant till completion of the anodization. After every hour, approximately 5 mL of fresh 4.8 M LiOH was added to the electrolysis vessel in order to compensate the loss occurred due to evaporation. After switching off the current the CE and WE were taken out of the electrolyte and rinsed intensively with tap water for 15 min and afterwards with deionized water for a further 10 min. Prior to the electrochemical characterization the samples were dried under air at 80 °C for 300 min and after the samples cooled down, the weight was determined upon a precise balance as described above. The sample preparation was repeated and 5 samples Ni42-Li127, 5 samples Ni42Li205 and 5 samples Ni42Li300 (Table 1) have been prepared this way.

## Results and Discussion

### *OER properties of untreated- and surface modified stainless steels*

We have examined 5 different steel specimens besides $IrO_2$-$RuO_2$ as potential OER electrocatalysts in $H_2SO_4$. Table 1 gives an overview of the samples/sample preparations and the corresponding OER key data.

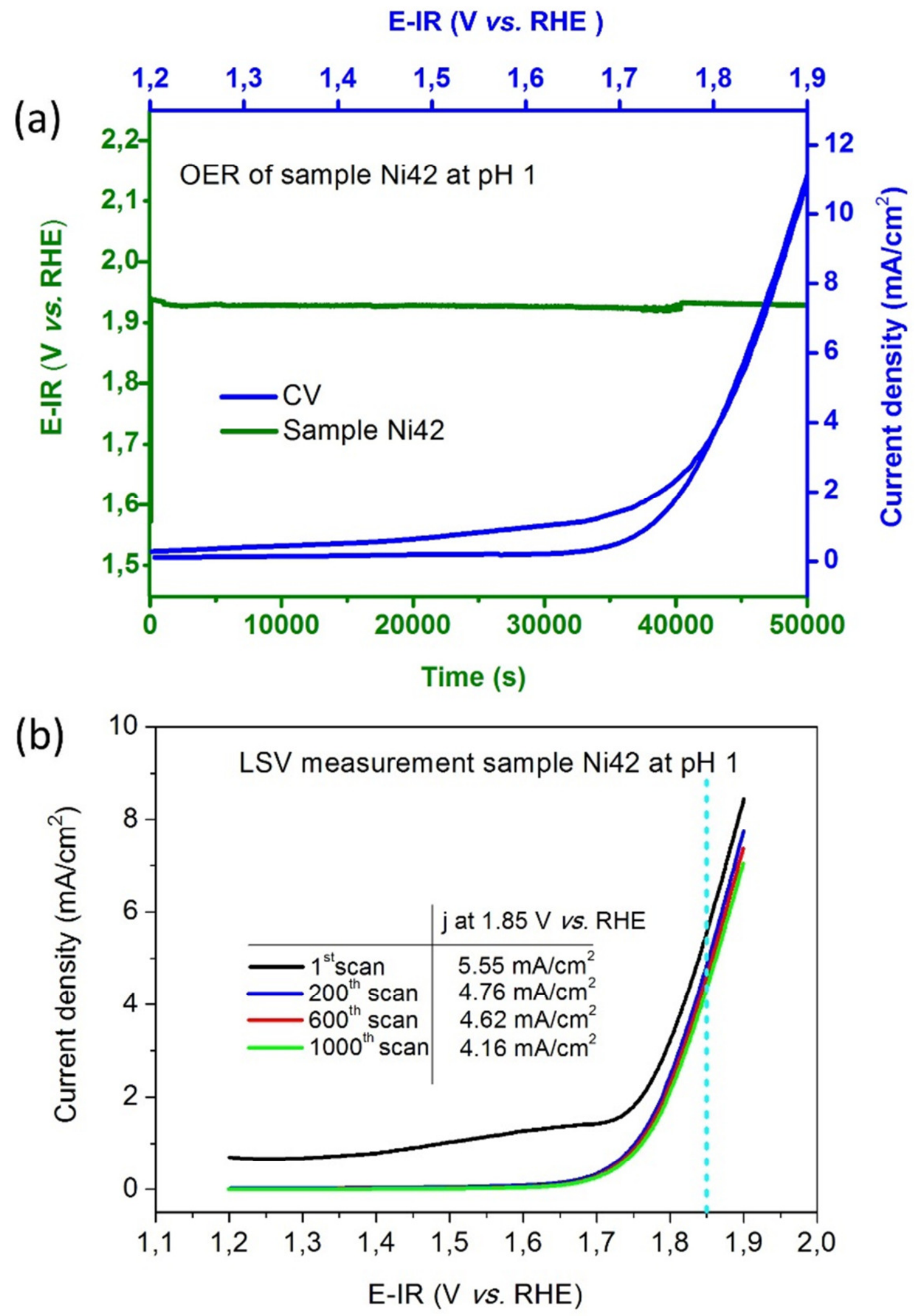

Figure 1. Steady state and non-steady state voltage current behavior of untreated steel Ni42 in 0.05 M $H_2SO_4$. CVs were recorded with a scan rate of 20 mV/s. Linear sweep voltammetry (LSV) was performed with a scan rate of 10 mV/s. Electrode area of all samples: 2 $cm^2$. Stirring of the electrolyte was performed for all measurements. **(a)** Cyclic voltammogram of sample Ni42; Long term chronopotentiometry plots of sample Ni42 at a current density of 10 $mA/cm^2$. **(b)** LSV measurements performed with sample Ni42 at 1.85 V *vs.* RHE.

| Sample name/<br><br>Material | Activation | | J max at 1.9 V *vs.* RHE derived from CV | Averaged potential (V *vs.* RHE) for 10 mA/cm$^2$ | Averaged weight loss (µg/mm$^2$) after 50 ks of chronopotentiometry at 10 mA/cm$^2$ | Resistivity $R_S/R_{CT}$ (Ω) at Offset potential ( V vs. RHE) derived from EIS |
|---|---|---|---|---|---|---|
| | Electroox. | Therm. | | | | |
| Ni42/AISI Ni42 | - | - | 11.3 | 1.935 | 185.4 | 4.8/ 10 (1.8); 5/ 2.5 (1.9) |
| Ni42-950/AISI Ni42 | | 950°C/ 720 min | 5 | 1.96 | 347.4 | — |
| Ni42Li127/AISI Ni42 | 4.8 m LiOH/127 min | | 15 | 1.85 | 48.7 | — |
| Ni42Li205/AISI Ni42 | 4.8 m LiOH/205 min | | 19 | 1.780 | 20.05 | 4.2/8 (1.8); 4.2/ 1.6 (1.9) |
| Ni42Li300/AISI Ni42 | 4.8 m LiOH/300 min | | 13.9 | 1.845 | 40.2 | — |
| $IrO_2$-$RuO_2$ | - | - | 32 | 1.710 | 8 | |

Table 1. Overview of the prepared samples (columns I), the performed surface modification (columns II, III) as well as the electrocatalytic properties of the samples (columns IV-VII).

Table 1 summarizes the electrocatalytic properties of untreated steel samples for the OER in 0.05 M $H_2SO_4$. As expected, the specimens consisting of untreated stainless steel Ni42 showed a significant increment of the current derived from the CV/LSV measurements due to onset of oxygen evolution (Sample Ni42; Figures 1a, 1b). Ni42 exhibited reasonable and stable potentials at constant current density under steady state conditions (Sample Ni42; Figure 1a; green curve).
The overpotential amounted to 707 mV (Sample Ni42) at 10 mA cm$^{-2}$ current density in 0.05 M $H_2SO_4$ (Figure 1a; green curve). However, the OER performance of sample Ni42 was found to be sensitive towards repeated dynamic variation of the voltage (Figure 1b) between 1.2 and 1.9 V *vs.* RHE: Thus *e.g.* the current density reached at 1.85 V *vs.* RHE decreased from 5.55 to 4.16 mA/cm$^2$ (sample Ni42, Figure 1b). However, the catalyst consisting of stainless steel Ni 42 is still on the level of some of recently developed OER electrocatalysts such as $Ba_2TbIrO_6$ [36], $MnCoTaSbO_x$ [48] and electrodes with 10 wt.% $Ir_{0.5}Ru_{0.5}O_2$ [37] . CoTiP was recently studied [49] in 0.5 M $H_2SO_4$ and was inferior ($\eta$= 971 mV at 8 mA/cm$^2$) to sample Ni42. In addition non-noble element containing compounds like CoP [31], NiB [31], CoB [31], NiMoFe [31], NiFe [31], $CoO_x$ [30],$NiFeO_x$[30], $NiO_x$[30] were found to be extremely instable towards OER in 1 M $H_2SO_4$ and less active with $\eta$~1000 mV at 10 mA/cm$^2$ [31].

Whereas the OER efficiency of Ni42 steel is still satisfying, a substantial dissolving of steel ingredients whilst long term OER in acids turned out to be an issue. Ni42 steel lost 185.4 µg per mm$^2$ upon OER polarization for 50 ks in 0.05 M $H_2SO_4$ performed at constant 10 mA cm$^{-2}$ (Table 1). This is a significant drawback ruling out the applicability of this material as an electrode in water electrolysis at low pH value. All the elements Ni42 steel is composed of could be determined via ICP-OES (Table S1) after 50 ks of OER in the electrolyte (0.05 M $H_2SO_4$) [20, 50]. The total amount of ions determined in the electrolyte is in good agreement with the mass deficit occurred to the samples through long term OER electrocatalysis (columns II and V of Table S1). As expected, Ni42 steel lost basically Fe and Ni besides small amounts of Mn.

Modeling of the frequency response (0.1 Hz-50 kHz) behavior of all steel samples discussed in this work at pH 1 at an offset potential that ensures oxygen evolution succeeded with the circuit of the so called Randles cell (Figure 2a) in which the double layer capacity is in parallel with the impedance due to the charge transfer reaction. Therefore, the Nyquist plot always shows a semicircle. The real axis value at the high frequency intercept can be interpreted as the solution resistance, the real axis value at the low frequency intercept can be interpreted as the sum of the solution and the charge transfer (CT) resistance ($R_{ct}$), respectively. The diameter of the semicircle therefore represents the CT resistance. The values for CT and solution resistance ($R_s$) of Ni42 can be taken from Table 1. The outcome (Figure 2b) is reasonable in light of the results gained from OER polarization experiments seen in Figure 1. At significant overpotential of 672 mV (1.9 V *vs.* RHE) Ni42 steel showed a CT resistance of 2,5 Ω. As expected, the radii of the corresponding circle in the Nyquist plot for one and the same sample decreases with increasing offset potential, which originates from an accelerated charge transfer (Figure 2b).

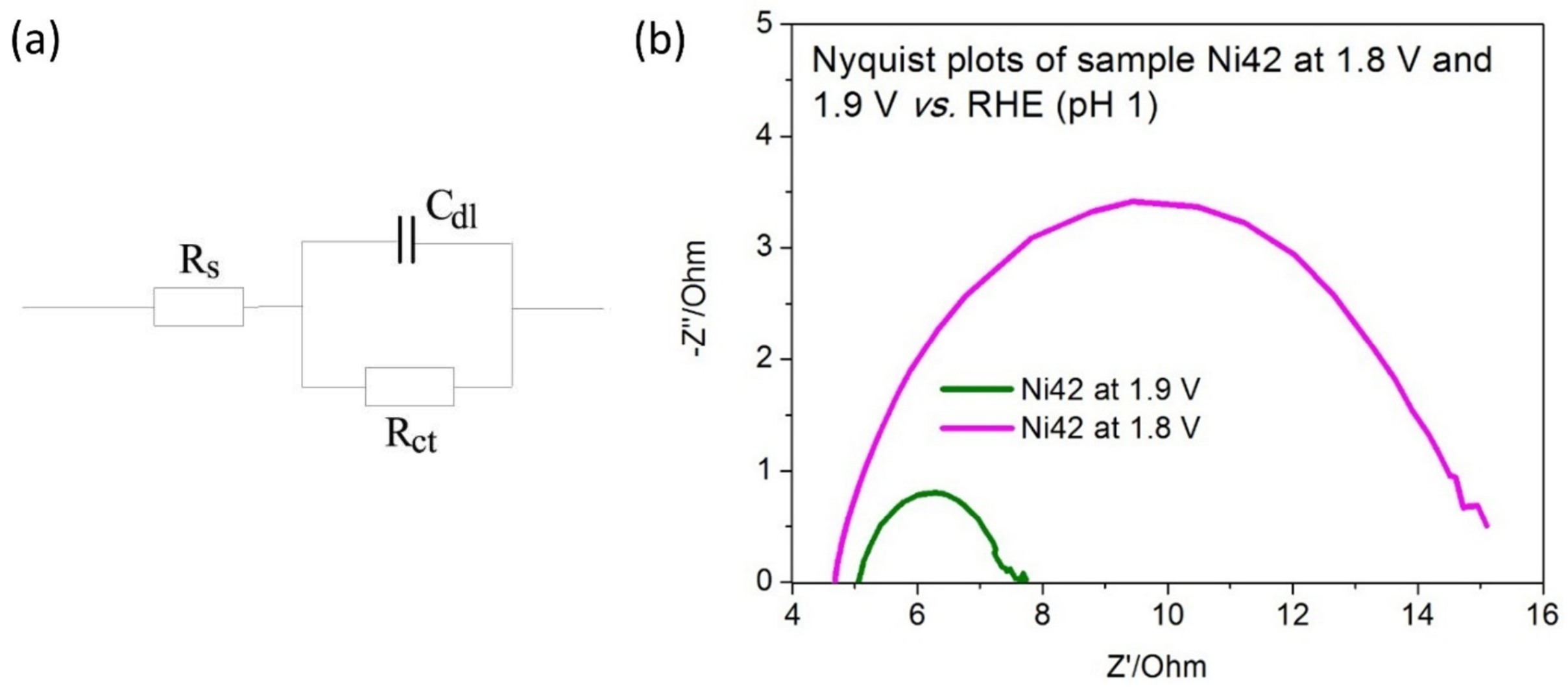

Figure 2. **(a)** The circuit of the Randles cell. **(b)** Nyquiest plots of the frequency response analysis of sample Ni42 at offset potentials of 1.8- and 1,9 V *vs.* RHE.

*OER properties of surface modified special alloy Ni42*

So far, surface modification procedures applied to Ni 42 steel which we have shown in previously published reports did not result in an electrode that durably produces oxygen in acidic regime [29]. Ni 42 steel electro-oxidized in 7.2 M NaOH for 300 min (sample Ni42-300 in our previous report) was found to be unstable towards OER at low pH values [29]. We modified the synthesis procedure, *i. e.* exchanged 7.2 M NaOH by 4.8 M LiOH and varied the duration (Sample Ni42Li127: 127 min; Sample Ni42Li205: 205 min; Sample Ni42Li300: 300 min). Notably: All Ni42XXX samples prepared this way exhibited a remarkably low weight loss (20.05-48.7 µg/mm$^2$; Table 1, Table S1) upon long term usage as OER electrode (50000 s of chronopotentiometry at j= 10 mA/cm$^2$) at pH 1. The mass deficit of Ni42Li205 amounted to (20.05 µg/mm$^2$) which represents a reduction by ~90% when compared with untreated Ni42 steel (Ni42: 185.4 µg/mm$^2$). For the ease of comparison, the OER key data of sample Ni42Li205 have been entered together with the ones of literature known compounds (Table 2). Degradation of electrocatalysts in acids when oxidative potentials are applied is well known, and we emphasize that even noble metal-based compounds such as $IrO_x/SrIrO_3$ show a certain mass deficit after long term polarization at positive potentials [51]. Also $MnO_x$ has been exploited as an OER electrocatalyst in different acidic regimes with pH value in between -0.5 and 2 [52]. However, current densities were found to be extremely low << 1mA/cm$^2$ (Table 2) at reasonable potentials and significant dissolution was found. Layered manganese-calcium oxide was investigated as prospective OER electrocatalyst in 0.1 M $HClO_4$ but dissolved even without oxidative potentials and exhibited poor OER efficiency [53]. All representatives of the sample series Ni42LiXXX are efficient and durable OER electrodes in acidic regime (Figures 3a-c, S1). The non-steady state- and steady state OER performance of samples Ni42Li127 and Ni42Li300 at pH 1 are similar (Figures 3a, b). Sample Ni42Li205 however showed the best overall OER properties (η= 552 mV; Ni42= 707 mV) derived from long term chronopotentiometry performed at j=10 mA/cm$^2$ (Figure 3b) and reached 16 mA/cm$^2$ at 1.9 V *vs.* RHE (Figure 3a; green curve). Even an extension of the chronopotentiometry measurement up to a duration of 150000 s did not change the OER performance (Figure 3c). Onset of OER in 0.05 M $H_2SO_4$ upon all surface modified Ni42 samples started at a potential as low as 1.68 V *vs.* RHE (Figure 3a). The onset for OER upon $IrO_2$ and $RuO_2$ in diluted sulfuric acid differs and amounted in between 1.45 and 1.53 V [37, 51]. Repeated LSV is a common tool to simulate fast aging of the electrode. The dynamic potential-current behavior of Ni42Li205 did not change substantially under repeated execution (Figure 4a, b). Under identical conditions (1.85 V *vs.* RHE) the current density drop after 1000 scans could be reduced from 1.39 mA/cm$^2$ (Ni42, Figure 1b) down to 0.81 mA/cm$^2$ (Ni42Li205, Figure 4a). Moreover, sample Ni42-Li205 exhibited after 2000 LSV scans an even slightly better OER performance (j=6,48 mA/cm$^2$ at 1.85 V vs. RHE) than after 1000 scans (j=6,12 mA/cm$^2$ at 1.85 V vs. RHE). A high durability of sample Ni42-Li205 towards electrocatalytic water oxidation can also be derived from cyclovoltammetry scans (Figure 4c). Even after 2500 scans the OER performance is still on the level of the start value (Figure 4c). To exclude that oxidation of the catalyst itself during OER experiments falsifies the evaluation of the OER performance it is recommended to quantify the charge to $O_2$ conversion rate. The Faradaic efficiency for the OER upon sample Ni42Li205 at pH 1 at 5 $mA/cm^2$ (10 $mA/cm^2$) amounted to 91.5% (79%) after 2000 s running time (Figures 5a, 5b). These are reasonable efficiencies for the OER upon a non-noble metal based electrocatalyst in highly corrosive media. A similar Ni based steel exhibited in 0.1 M KOH at 10 $mA/cm^2$ 75.5% charge to oxygen conversion[41] after 4000 s. Multiple point nitrogen gas adsorption BET measurements (Figure S2) were carried out in order to determine the change in specific surface area through electro-oxidation. In our recent investigations, electro-oxidation of stainless steels in alkaline media did not lead to substantial changes in the size of the specific surface [29, 20]. In comparison with untreated Ni42 steel (sample Ni42), which surface area was recently determined upon BET measurements and amounted to 0.354 $m^2/g$ [29], the surface area of Ni42Li205 was proved slightly higher (0.52 $m^2/g$; Figure S2). This agrees well with the outcome from SEM experiments carried out with sample Ni42/Ni42Li205 (SEM top view images Figures 6a-d) that uncover some particles on the surface of Ni42Li205 increasing the surface area.

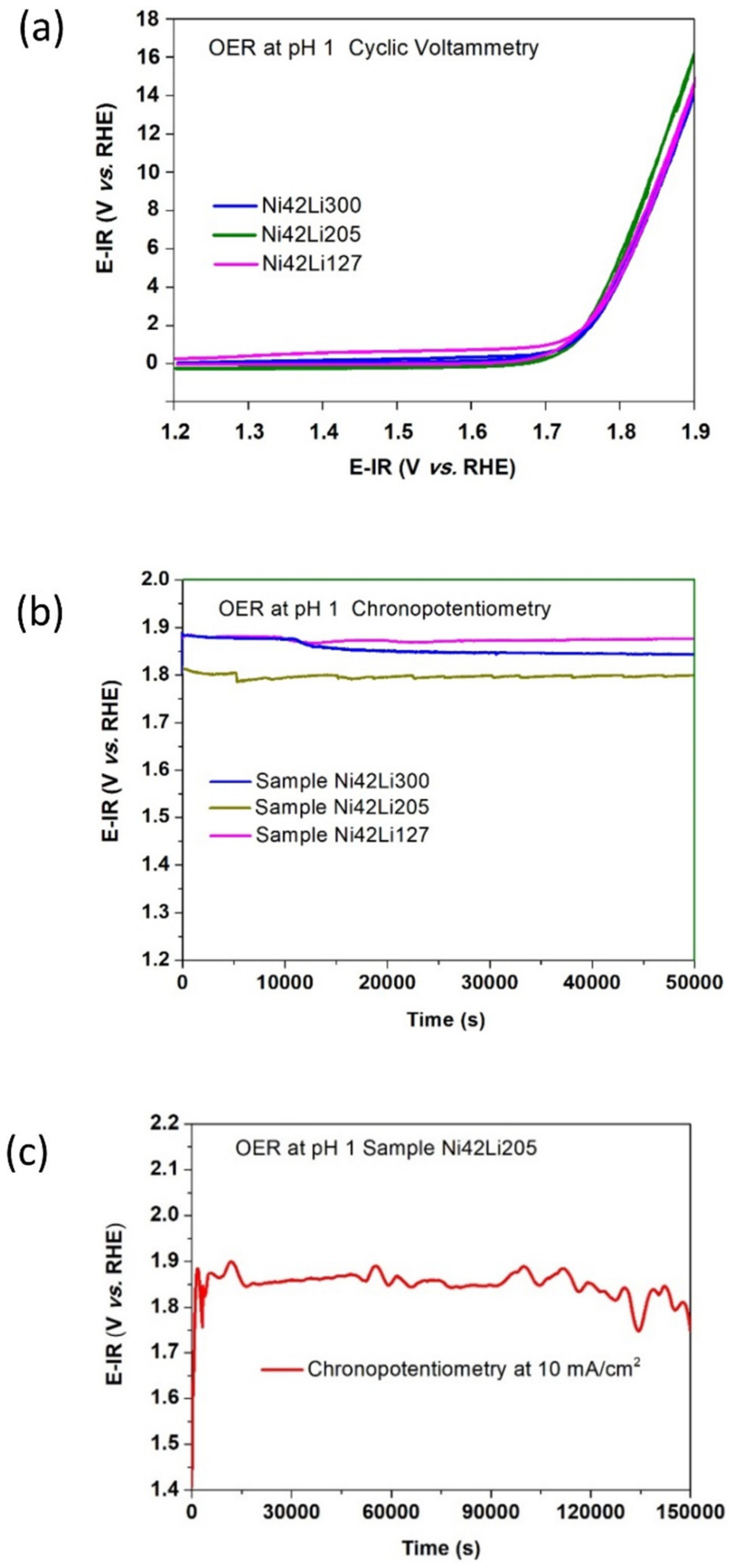

**Figure 3**. The OER properties of surface oxidized 1.3917 steel in 0.05 M $H_2SO_4$. CVs were recorded with a scan rate of 20 mV/s. Electrode area of all samples: 2 $cm^2$. Stirring of the electrolyte was performed for all measurements. Chronopotentiometry measurements were performed at 10 $mA/cm^2$ current density. **(a)** Cyclic voltammogram and chronopotentiometry plot of sample Ni42Li127. **(b)** Cyclic voltammogram and chronopotentiometry plot of sample Ni42Li205. (c) Chronopotentiometry plot (derived from long-term measurement) of sample Ni42-Li205**. (d)** Cyclic voltammogram and chronopotentiometry plot of sample Ni42Li300.

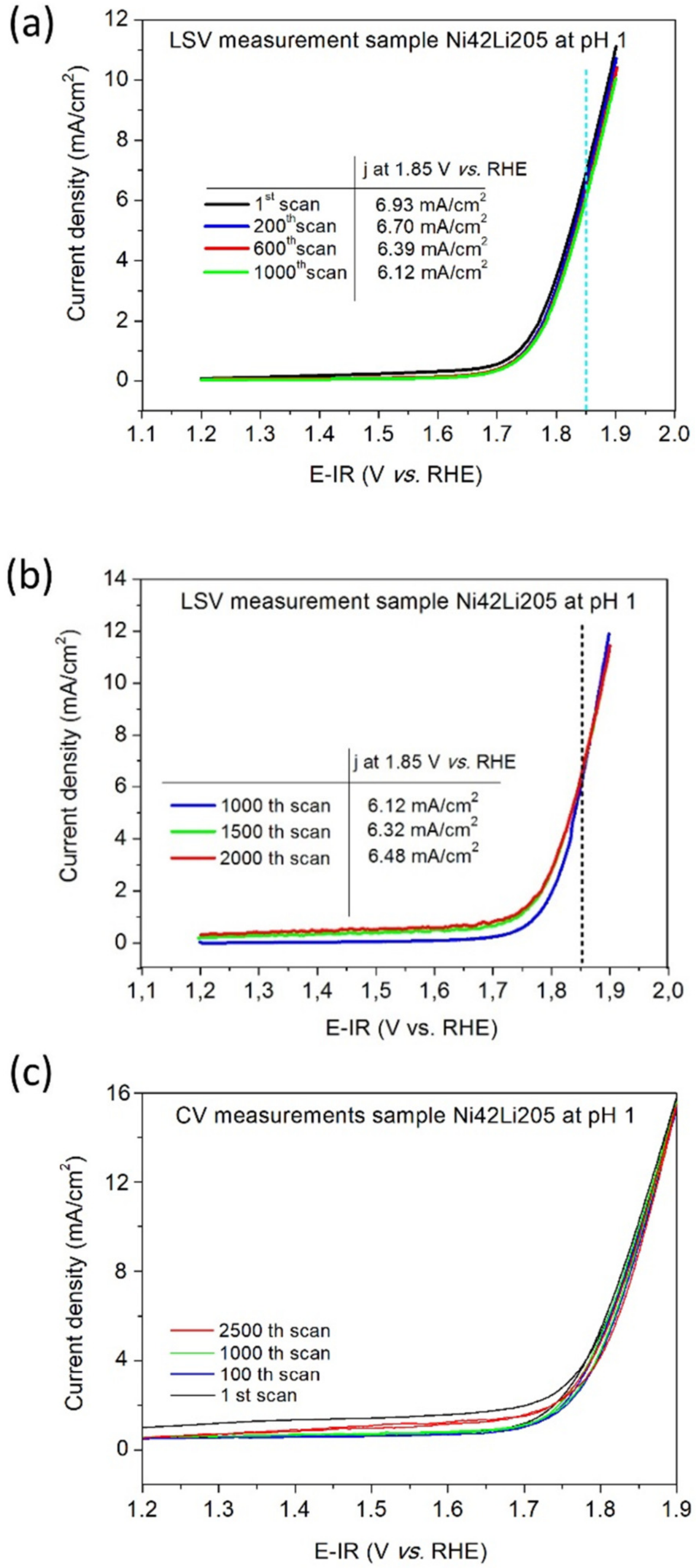

**Figure 4**.OER properties of sample Ni42-Li205 upon steady state and non-steady state conditions. Electrode area of the sample: 2 $cm^2$. The scan rate was adjusted to 20 mV/sec for the CV measurements, 10 mV/sec for the LSV measurements respectively. Repeated LSV scans of sample Ni42Li205 in between 1.2-and 1.9 V *vs.* RHE at pH 1; $1^{st}$- 1000 th scan **(a)**; 1000 th-2000 th scan **(b)**. Repeated CV scans of sample Ni42Li205 in between 1.2-and 1.9 V *vs.* RHE at pH 1 **(c).**

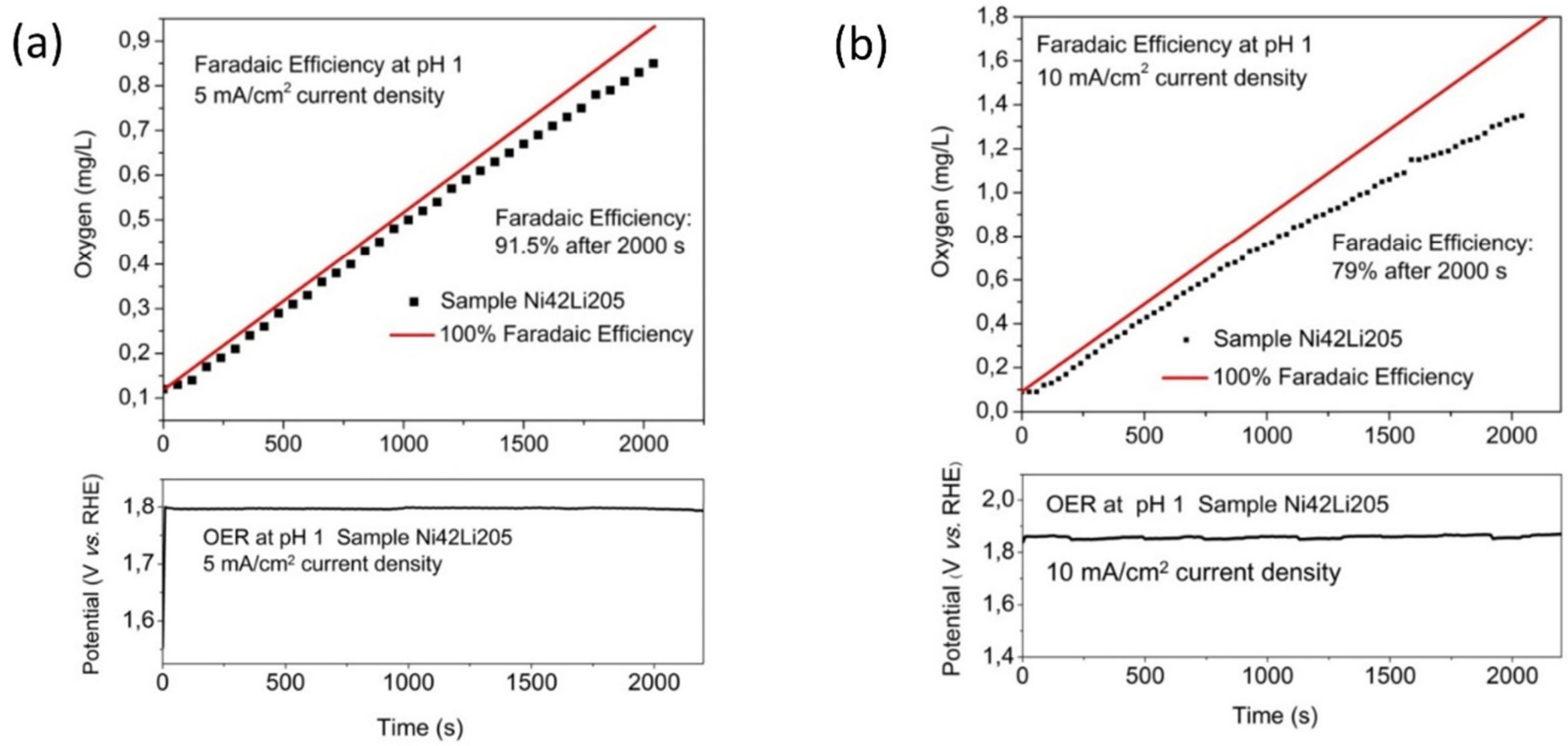

Figure 5**(a) top:** Correlation of oxygen evolution upon sample Ni42Li205 in 0.05 M $H_2SO_4$ (dotted curve) with the charge passed through the electrode system (red line corresponds to 100% Faradaic efficiency). Electrode area of the sample: 2 $cm^2$. Faradaic efficiency of sample Ni42Li205 at 10 mA $cm^{-2}$ current density; Total volume= 2080 mL, Faradaic efficiency after 2000 s: 79%; line equation: Y=0.00079 * X +0.09, where Y represents the oxygen content (mg/L) and X represents the time (s); **bottom**: Corresponding chronopotentiometry plot. **(b) top**: Faradaic efficiency of sample Ni42Li205 at 5 mA $cm^{-2}$ current density; total volume= 2080 mL, Faradaic efficiency after 2000 s: 91.5%; line equation: Y=0.00039 * X, where Y represents the oxygen content (mg/L) and X represents the time (s); **bottom:** Corresponding chronopotentiometry plot.

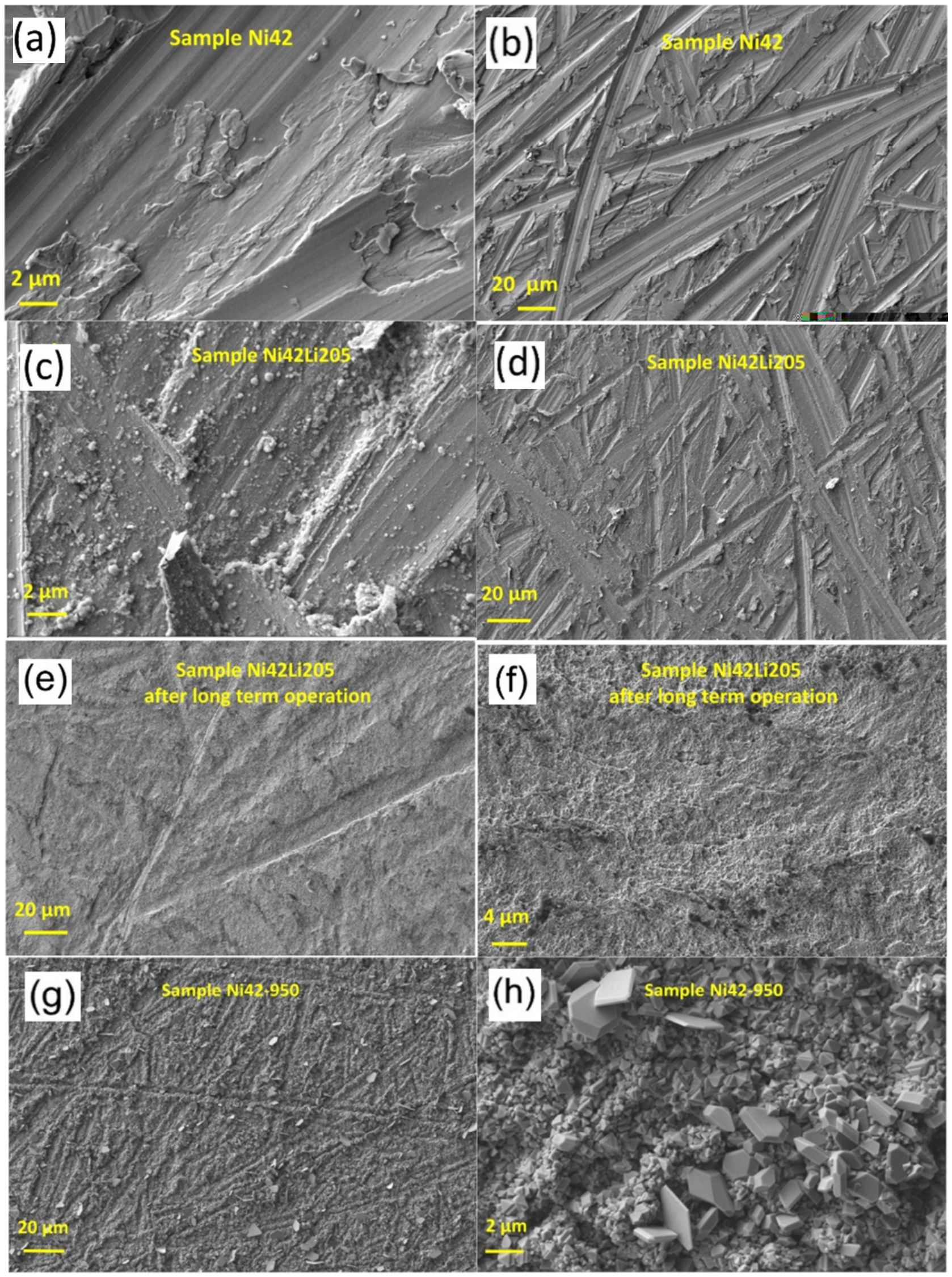

**Figure 6**. SEM images of untreated stainless steels (a, b) as well as surface oxidized stainless steels (c, d, e, f, g, h). Accelerating voltage: 5 kV; detector: secondary electron detector.

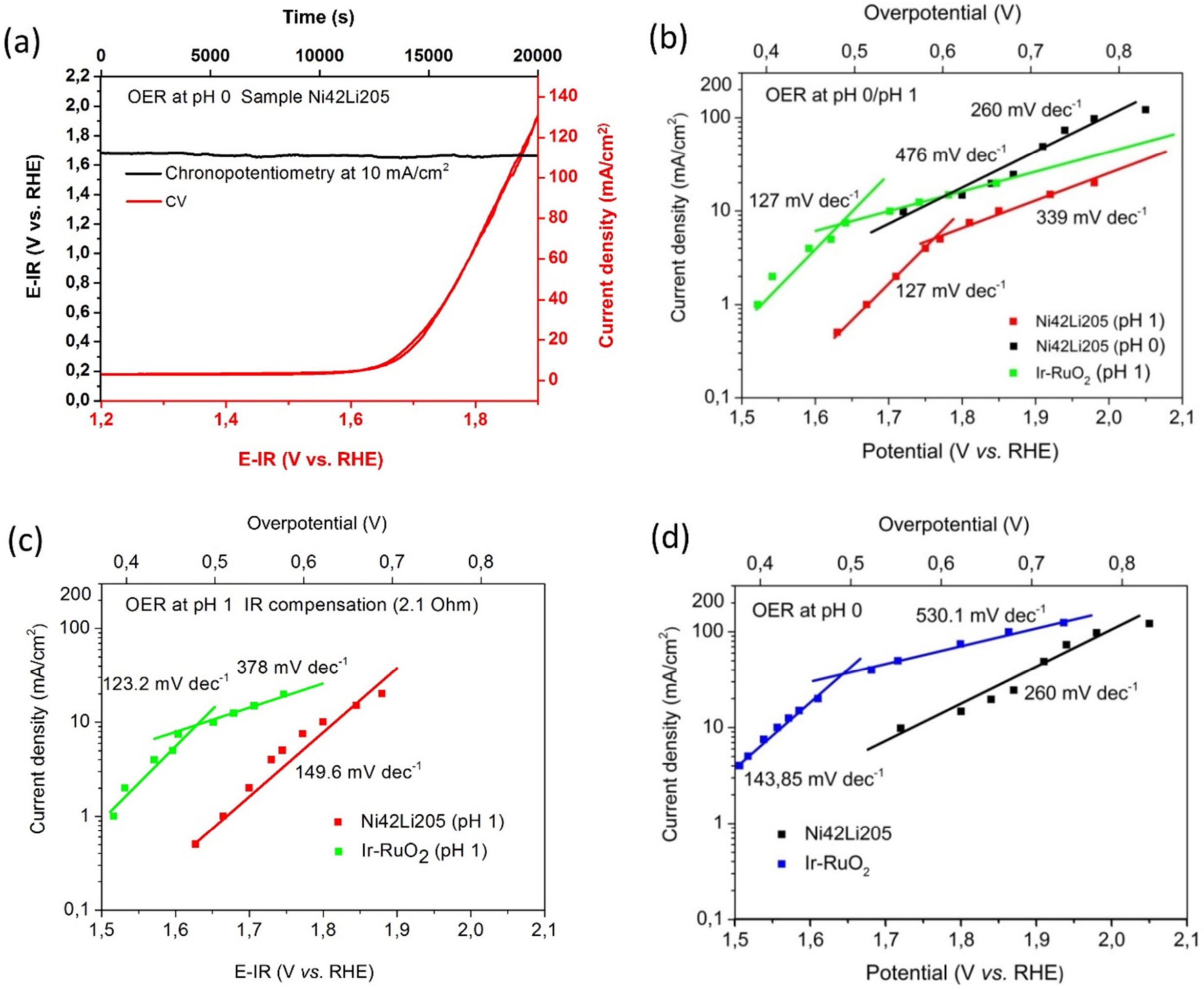

Figure **7.** The OER properties of surface oxidized Ni42 steel and $RuO_2/IrO_2$ in 0.05 M $H_2SO_4$ and 0.5 M $H_2SO_4$. CVs were recorded with a scan rate of 20mV/s. Electrode area of all samples: 2 $cm^2$. Stirring of the electrolyte was performed for all measurements. Chronopotentiometry measurements were performed at 10 $mA/cm^2$ current density. Data used for Tafel plots are based on averaged values derived from 200 s chronopotentiometry scans. **(a)** Cyclic voltammogram and chronopotentiometry plot of sample Ni42Li205 determined at pH 0. **(b)** Tafel plots of samples Ni42Li205 (pH 1 and pH 0) and Ir-$RuO_2$ (pH 1) without correction of the voltage drop. **(c)** Tafel plots of samples Ni42Li205 and Ir-$RuO_2$ at pH 1 with correction of the voltage drop on the basis of 50% correction of the solution resistance. **(d)** Tafel plots of samples Ni42Li205 and Ir-$RuO_2$ at pH 0 without correction of the voltage drop.

The topography of samples Ni42 and Ni42 Li205 in micrometer scale as seen by using electron microscopy (Figure 6a, 6c) have been fully confirmed upon atomic force microscopy (Figures S3). Longer usage of Ni42Li205 as OER electrode did not significantly change the topography of the periphery (Figures 6e, f).

We determined polarization data of sample Ni42Li205 also at pH 0 (0.5 M $H_2SO_4$; Figures 7) to make it easier to compare our results with earlier published ones. The substantial stronger current-voltage ratio (in comparison with the pH 1 data) can be clearly taken from the cyclic voltammogram (Figure 7a) and the chronopotentiometry plot ($\eta$= 445 mV for 10 mA $cm^{-2}$ (Figure 7a). This value underpins the high efficiency of the electrocatalyst and demonstrates the competitiveness of sample Ni42Li205 to recently developed electrocatalysts. The best OER performance achieved in acids at pH 0 upon a noble metal containing catalyst ($IrO_x$/ /$SrIrO_3$) was shown by Seitz *et. al.* *(*$\eta$= 280 mV at 10 mA/ $cm^2$) [54]. However, also this material was not 100% stable at oxidative potentials: Sr was determined via ICP-OES in the electrolyte used for long term polarization experiments [54]. A F-doped CuMn-oxide based OER- and oxygen reduction electrode intended to be suitable for electrocatalysis in sulfuric acid was recently shown [55]. Unfortunately, non-IR corrected data plus a detailed evaluation of the mass loss whilst usage have not been shown. In addition, the OER activity when derived from chronopotentiometry data was mediocre ($\eta$= 320 mV at ~ 1.5 mA/$cm^2$ in 0.5 M $H_2SO_4$). Frydendal reported on Ti stabilized $MnO_2$ as prospective OER electrocatalyst and determined $\eta \sim 550$ mV at 3 mA/$cm^2$ in 0.05 M $H_2SO_4$ [56].

Tafel slopes of samples Ni42Li205 have been determined in 0.05 M and 0.5 M $H_2SO_4$ and amounted at higher potentials to 260 mV $dec^{-1}$ (pH 0) and 339 mV $dec^{-1}$(pH 1), at lower potentials to 127 mV $dec^{-1}$ (pH 1), respectively (Figure 7b). Electrodeposited Co-oxide exhibited a slope of 529 mV $dec^{-1}$ for OER in 1 M $BF_4^-$ at pH 1 [57].

Without IR compensation of the data, dual Tafel behavior was found for Ni42Li205 at pH 1 (Figure 7b) and single Tafel behavior was obtained along the entire potential range at pH 0 (Figure 7b). Voltage drop compensation realized by 50 % compensation of the solution resistance of 4.2 Ohm (Table 1) becomes more dominant at higher current densities, resulted in a quenching of the dual Tafel behavior (Figure 7c) at pH 1 and leads to a Tafel slope of 149.6 mV $dec^{-1}$ (Figure 7c). Noble metal containing catalysts [58, 59] ,especially $IrO_2$-$RuO_2$ [30, 60] are known for its high OER efficiency in acidic regime. Commercially available $IrO_2$-$RuO_2$ (sample Ir-$RuO_2$) outperforms sample Ni42Li205, as can be taken from the Tafel lines of Ir-$RuO_2$ which were found to be shifted to lower overpotentials by ~ 100 mV at pH 1, by ~ 180 mV at pH 0 respectively (Figure 7c, d). Recently, an overpotential in-between 0.27 and 0.32 V for OER upon $IrO_2$ at 10 mA $cm^{-2}$ in 1 M $H_2SO_4$ [30] and ~ 420 mV in 0.1 M $HClO_4$ [60] was determined and agrees quite well with our values derived from Tafel lines recorded for sample Ir-$RuO_2$ in 0.5 M $H_2SO_4$ (Figure 7d). Sample Ir-$RuO_2$ showed in pH 1 medium the same Tafel slope as sample Ni42Li205 did but at lower potential region (127 mV $dec^{-1}$, Figure 7b). The dual Tafel characteristics found for Ir-$RuO_2$ at pH 1 (Figure 7b) when non-IR corrected data have been analyzed were not quenched upon performing ohmic correction of the voltage drop (Figure 7c). This can be explained rationally by a transition between different OER mechanisms or by different rate-determining steps on moving from low- to high-overpotential regions. Dual Tafel behavior with a transition from low to high Tafel slope at 1.08 V *vs.* SCE was obtained for $RuO_2$ by Lyons et al. [61]

The difference in OER performance of both samples (Ni42Li205 and Ir-$RuO_2$) substantially decreases with increasing potential, as can be taken from the corresponding Tafel lines, which move towards each other at higher potentials (Figures 7b-7d). At pH 0, sample Ir-$RuO_2$ exhibited Dual Tafel characteristics with slopes of 143.85 and 530.1 mV $dec^{-1}$ at lower, higher potential, respectively (Figure 7d). The difference in OER performance again decreases with increasing potential and was found to be reduced to almost zero at a potential of ~ 2 V *vs.* RHE (Figure 7d). Thus, to sum up, the advantage of $IrO_2$-$RuO_2$ over sample Ni42Li205 with respect to current voltage behavior was not as substantial as expected, and moreover even $RuO_2$-$IrO_2$ shows a “bleeding effect” (8 µg/$mm^2$; Table 1) when used as an OER electrode in acidic regime [32, 62].

*The origin of the OER properties of Ni42Li205*

Figure 8 displays Nyquist plots of the samples Ni42 and Ni42-Li205 recorded at pH 1. A lower CT resistance was determined at both offset potentials (1.8 V *vs.* RHE and 1.9 V *vs.* RHE) for sample Ni42-Li205 (1.6 Ω/1.9 V; 8 Ω/1.8 V) when compared to sample Ni42 (2.5 Ω/1.9 V; 10 Ω/1.8 V). On the basis of these findings it is reasonable to assume that the oxide based periphery of sample Ni42Li205 caused upon electro oxidation in LiOH is responsible for the reduced CT resistance and in turn reduces the voltage drop through the outer zone whilst OER in acidic regime. The outcome from frequency response analysis is therefore in good agreement with the lower overpotential of sample Ni42-Li205 derived from polarization experiments when compared to the one of sample Ni42 (Figures 1a, 3b and 4).

In recent years, the Tafel slope derived from OER polarization tests has been exploited as a descriptor for the stability of materials towards OER [63]. This concept is based on two different routes for the origin of oxygen formation whilst water splitting. The so called “oxide route” is used for materials that releases oxygen out of the metal-oxide containing surface [64, 65] whereas for a different group of materials adsorbed water molecules represent the oxygen source responsible for the OER (solution route) [66]. Typically, the oxide route leads to a more dominant dissolution process upon disruption of the surface, *i.e.* represents the source for instability. In addition, these materials exhibit significantly lower Tafel slopes than the "solution route materials" [63]. Notably: Sample Ni42Li205 exhibited a quite high Tafel slope of 149.6 mV/dec whereas untreated Ni42 steel showed a substantially lower Tafel slope value of 87.6 mV/dec (Figure S4). It therefore seems to be reasonable to assume that electro oxidation of Ni42 steel in LiOH (sample Ni42Li205) results in the formation of a metal oxide containing outer zone that supports solution route based oxygen evolution in acidic regime accompanied by a good stability of the catalyst (Figures 3b, 3c and 4) rather than the release of oxygen out of the oxide containing outer sphere itself. In contrast to this finding, non-pre-oxidized Ni42 steel forms an oxide-based periphery during the electrochemical measurements in acidic regime that supports oxide route based oxygen evolution in acids finally resulting in less pronounced stability (Figure 1b). Moreover, sample Ni42Li205 and sample Ir-$RuO_2$ exhibited similar Tafel slopes in a wide voltage range (Figure 7b) that might be interpreted as a hint for satisfying stability in both cases (Table 1).

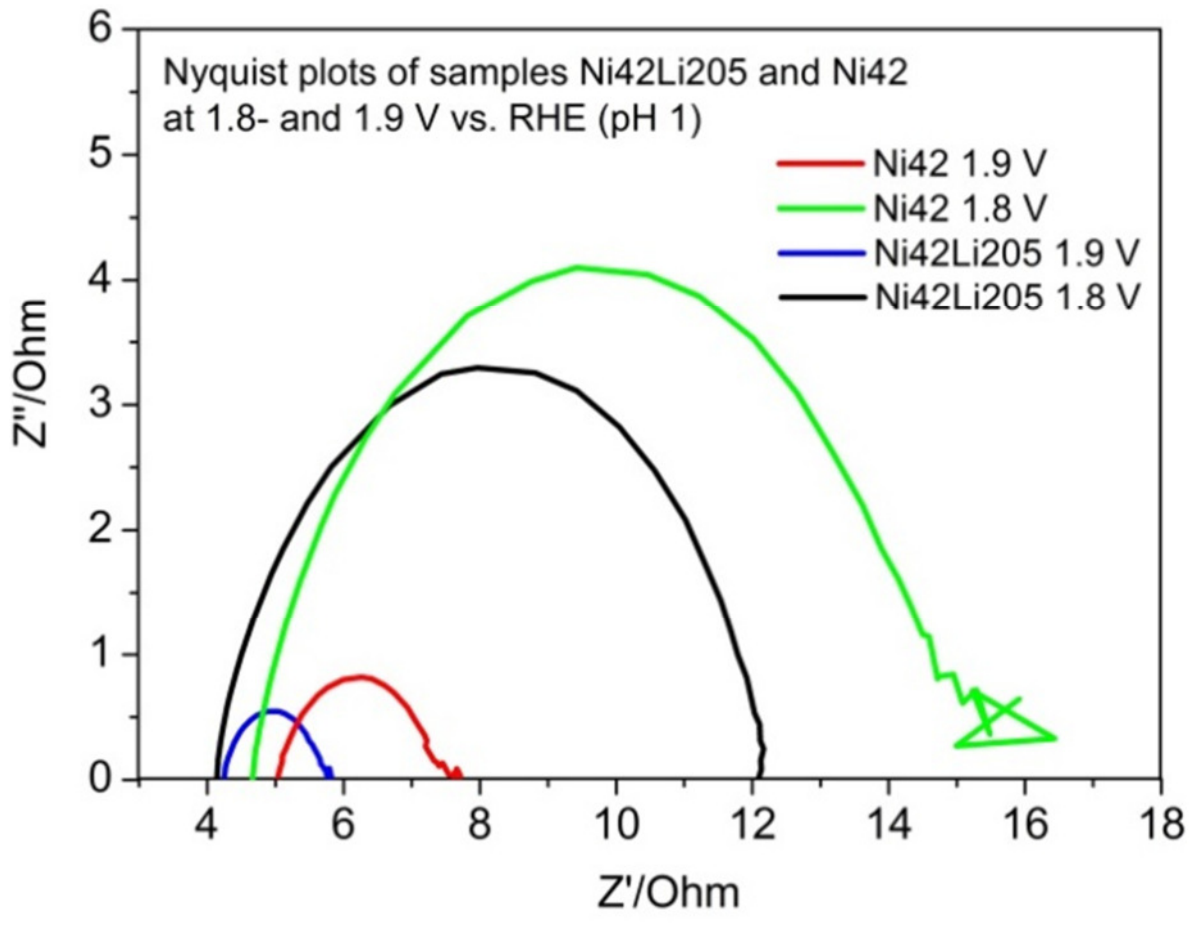

**Figure 8**. Nyquist plots of samples Ni42 and Ni42Li205 recorded in 0.05 M $H_2SO_4$. The offset potential was set to 1.8 and 1.9 V *vs.* RHE.

The XRD results (Figure S5) determined before and after carrying out OER tests do not indicate crystalline oxides on the surface of Ni42Li205. This was further confirmed by SEM top view images (Figures 6 c-f). The XRD measurements were carried out in grazing incidence mode, that is, they were sensitive to the surface (Figure S5). While using this technique we avoided that the beam permeates through the "surface layer" and detects the matrix. The microstructure has been investigated via FIB-SEM investigations (Figure 9) before and after OER testing. No differences regarding the microstructure of the outer sphere can be derived from cross section images of Ni42/Ni42Li205 achieved from FIB-SEM investigations (Figure 9 a, b) revealing that sample Ni42Li205 does not have a classical substrate-layer architecture known from samples achieved from electro deposition techniques. As shown in our previous report [20], this can be additionally seen as a source of stability whilst electrocatalytically initiated long term OER.

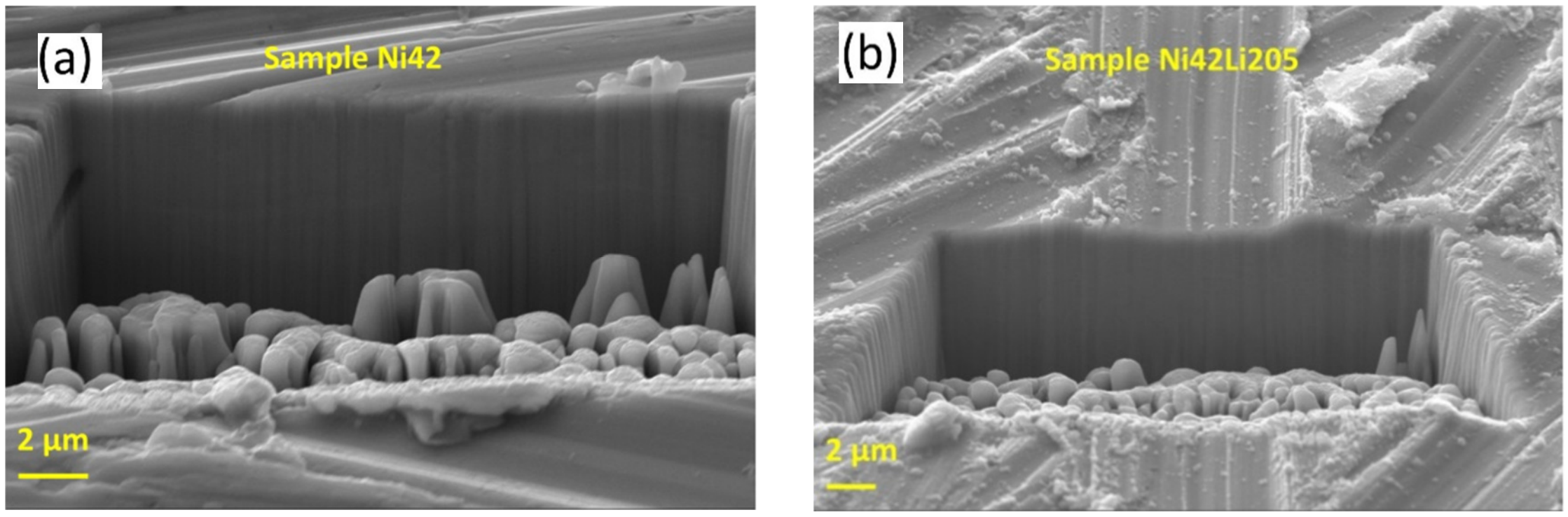

**Figure 9**. Cross sectional analysis of untreated- and surface modified steel derived from FIB-SEM experiments. The accelerating voltage was adjusted to 20 kV and the SEM images were acquired using a backscattered detector (a), a secondary electron detector (b) respectively. Ga beam settings: 2 nA, 30 kV.

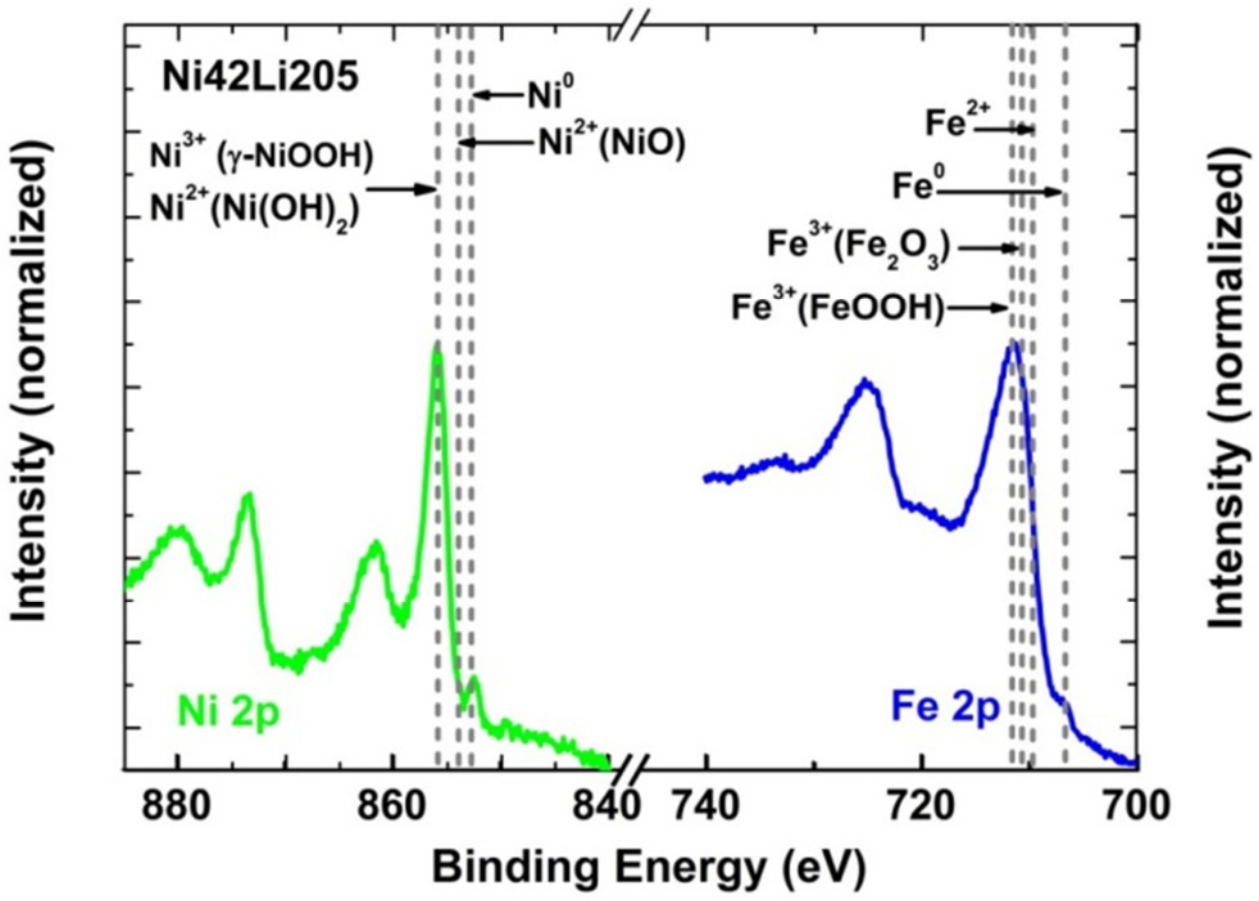

**Figure 10.** High resolution XPS (Fe 2p- and Ni 2p core level) spectra of sample Ni42Li205. Binding energies of reference compounds are indicated by vertical lines as a guide to the eyes.

XPS spectra of sample Ni42Li205 recorded after 4000 s of OER can be taken from Figure 10. The corresponding spectrum of untreated Ni42 steel has been shown in our previous report [29]. The $2p_{3/2}$ positions of some reference compounds[67, 68, 69, 70, 71] are indicated by gray vertical bars. The composition of the periphery of Ni42Li205 derived from the cationic distribution (31.89% Ni, 68.11% Fe) confirmed our expectation that a Ni containing compound is the predominant cationic species of the surface oxidized Ni42 alloy (Table S2). In comparison with sample Ni42 the Ni content was increased from 26.6 to 31.89 at%, whereas the iron content was found to be decreased from 72,4% to 68.11% (Table S2). Due to the lack of mass loss during electro-activation (Table S3) and the absence of hints that support the hypothesis that dissolution at least of some ingredients of the steel takes place during activation in LiOH (no coloration of the electrolyte), we conclude that electro migration of Ni caused by a momentum transfer e-→ M+ at high current densities [72] to the surface is the most likely origin for the Ni enrichment in the outer sphere of the activated steel obtained during anodization [29]. Earlier contributions [29, 41, 73] related to OER upon Ni containing films in alkaline media stated that γ–NiO(OH) is most likely the catalytic active species. Our XPS findings also unmask γ–NiO(OH) as the driving force for OER on the surface of Ni42Li205 in acidic media (Figure 10). In addition, small amounts of metallic Ni and Fe (See peaks located at 852.5 eV and 706.7 eV; Figure 10, right side) were detected. The presence of $Fe^{2+}$ can be excluded given the binding energies of the $Fe2p_{3/2}$ core level spectra. The FeOOH species of iron dominates rather than $Fe_2O_3$ on the surfaces of Ni42Li205 (Figure 10 right side), which agrees very well with a recent *operando* X-ray absorption spectroscopy (XAS) study, in which (Ni, Fe)OOH catalysts were characterized during OER operating conditions [74].

Besides activation upon electro-oxidation in aqueous solution at moderate temperatures (leading to samples Ni42Li127, Ni42Li205 and Ni42Li300) also thermal treatment under air has been applied to Ni42 steel resulting in sample Ni42-950 (12 h heat treatment at 950 °C). The basic idea was to generate a crystalline metal oxide based periphery whereas aqueous solution based approaches carried out at room temperature are known to result in amorphous metal oxides[43]. SEM experiments indeed exhibited crystal formation upon high temperature (Figures 6 g,h). We expected that crystalline oxides show a lower solubility in acids and should therefore end up in lower weight loss whilst chronopotentiometry experiments as known from literature [32, 75]. This expectation was not met. As summarized in Table 1 all electro-oxidized samples (Ni42Li127, Ni42Li205 and Ni42Li300) showed a substantially reduced weight loss upon long term usage as OER electrocatalyst when compaed with untreated alloy Ni42 (Table 1). Sample Ni42-950 however, exhibited after long term OER polarization in 0.05 M $H_2SO_4$ almost twice the mass deficit (347.4 µg/mm$^2$) sample Ni42 showed (185.4 µg/mm$^2$, Table 1). The layer (visible by eye) grown on the matrix upon heating to 950 °C consisted of Ni-Fe-oxide (Table S1) and was found to be diluted whilst long term chronopotentiometry. In addition to disappointing OER stability characteristics of sample Ni42-950 the potential/current behavior derived from polarization measurements carried out in acids under steady state or non-steady state conditions was rather weak (See Table 1).

| Sample | Average Overpotential (mV) at 10 mA/cm$^2$ (pH) | Tafel slope (pH) | Faradaic Efficiency (mA/cm$^2$) | Averaged weight loss (µg/mm$^2$) after 50 ks of chronopotentiometry in 0.05 M $H_2SO_4$ at 10 mA/cm$^2$ | Ref. |
|---|---|---|---|---|---|
| Surface modified Steel (Ni42Li205) | 552 (1)<br>445 (0) | 127 mV dec$^{-1}$ (1)<br>260 mV dec$^{-1}$ (0) | 79% (10 mA/cm$^2$)<br>91.5% (5 mA/cm$^2$) | 20.05 | This work |
| $IrO_x/SrIrO_3$ | 280 (0) | ~39 mV dec$^{-1}$ | Was not quantified | - | 54 |
| $Cu_{1.5}Mn_{1.5}O_4$ | 322 (0) at 1.5 mA/cm$^2$ | 65.7 mV dec$^{-1}$ | ~ 100%<br>(1.5 mA/cm$^2$) | - | 55 |
| $CoO_x$ | ~400 (3.4) at 1 mA/cm$^2$ | 529 mV dec$^{-1}$ (0) | ~ 95% at pH 3.4<br>(?? mA/cm$^2$) | - | 57 |
| $RuO_2$-$IrO_2$ (Ir-$RuO_2$) PVD sputtered on Titania | 522 (1)<br>340 (0) | 127 mV dec$^{-1}$ (1)<br>143.85 mV dec$^{-1}$ (0) | - | 8 | This work |
| $MnO_x$ | 219 (1) at 1µA/cm$^2$ | 127 mV dec$^{-1}$ (2.5) | ~ 100% at pH 2.5<br>(0.3 mA/cm$^2$) | - | 52 |
| $IrO_2$ on glassy carbon | 330 (0) at 15 mA/cm$^2$ | - | - | - | 76 |
| $IrO_2$-$RuO_2$ on Sb doped $SnO_2$ NP | 260 (0) at 1 mA/cm$^2$ | 60 mV dec$^{-1}$ (0) | - | - | 77 |
| (Mn-Co-Ta-Sb)$O_x$ | ~550 (-0.3) at 3 mA/cm$^2$ | - | - | - | 48 |
| $IrO_2/Nb_{0.05}Ti_{0.95}O_2$ | ~370 (0)<br>at 2 mA/cm$^2$ | 282 mV dec$^{-1}$ (0) | - | - | 78 |
| $CeO_2$ doped $RuO_2$-$IrO_2$ | 272 (0) at 4.8 mA/cm$^2$<br>(Based on CV data!) | 47 mV dec$^{-1}$ (0) | - | - | 34 |
| CoTiP | 972 (0) at 8 mA/cm$^2$ | - | - | - | 49 |
| Ti-stabilized $MnO_2$ | 547 (1) at 3 mA/cm$^2$ | 170 mV dec$^{-1}$ (1) | - | - | 56 |
| $RuO_2$ | 220 (1) | - | 90% at 5 mA/cm$^2$ | Significant dissolution | 32 |
| $IrO_2$ | 380(1) | - | 99% at 5 mA/cm$^2$ | Significant dissolution | 32 |
| Ru | 340 (-0.3)<br>- | - | 92% at 10 mA/cm$^2$ | 8 | 31 |
| Ir | 340 (-0.3) | - | 93% at at 10 mA/cm$^2$ | - | 31 |

**Table 2**. The OER key data of recently developed electrocatalysts determined at low pH.

## Conclusion

All known electrocatalysts that durably and efficiently convert electricity at low pH values into oxygen, which is a requirement for the exploitation as electrodes in (PEM) electolyzers, contain noble metals at least in the form of additives. This work evaluates the suitability of untreated Ni42 steel as well as surface modified Ni42 steel for usage as oxygen evolving electrodes in acids with pH $\leq$1. The Ni42 steel was, albeit not competitive to up to date OER electrocatalysts, found to be reasonably active. However, it proved to be unstable whilst electrocatalytically initiated oxygen evolution in acidic media. As a matter of fact, it exhibited a substantial loss of ingredients when oxidative potentials were applied in 0.05 M sulfuric acid. A heat treatment of the material at 950 °C created crystalline oxides on the periphery but however was found to be unsuitable to stabilize it towards positive potentials in acid. In addition, this surface modification procedure led to unsatisfying overall OER properties.

Ni42 steel, electro-oxidized in LiOH, represents the best outcome of this study. The sample was found to be highly efficient ($\eta$= 445 mV at 10 mA $cm^{-2}$ at pH 0) and reasonable stable (weight loss: 20 µg/$mm^2$ after 50 ks of chronopotentiometry at pH 1) towards OER in $H_2SO_4$. We assume that electro oxidation of Ni42 steel in LiOH (sample Ni42Li205) results in the formation of a metal oxide containing outer zone that supports solution route-based oxygen evolution in acidic regime accompanied by a good stability of the catalyst. To the best of our knowledge, a similar activity and durability for OER in acids proven by comprehensive testing has not been shown for a catalyst solely consisting of cheap abundant elements.

## Conflicts of interest

There are no conflicts to declare

## Supporting Information

Electronic Supplementary Information (ESI) containing a description of sample preparations, a description of all measurements as well as additional Figures and tables is available:

**Acknowledgements**: HS, MS, PH and WH were supported by the European Research Council (ERC-CoG-2014; project 646742 INCANA). The authors thank the German Research Foundation for funding of a focused ion beam unit (INST 190/164-1 FUGG).

**Conflict of Interest Disclosure**: The authors declare no competing financial interest.

[37] N. Baumann, C. Cremers, K. Pinkwart, J. Tübke, *Fuel Cells,* 2017, **17**, 259-267.
[38] D. Lebedev, M. Povia, K. Waltar, P. M. Abdala, I. E. Castelli, E. Fabbri, M. V. Blanco, A. Fedorov, C. Coperet, N. Marzari, T. J. Schmidt, *Chem. Mater.,* 2017, **29**, 5182-5191.
[39] E. Oakton, D. Lebedev, M. Povia, D. F. Abbott, E. Fabbri, A. Fedorov, M. Nachtegaal, C. Coperet, T. J. Schmidt, *ACS Catal.*, 2017, **7**, 2346-2352.
[40] J. W. Medlin, M. M. Montemore, *Nature Chemistry* 2015, **7**, 378-380.
[41] H. Schäfer, S. Sadaf, L. Walder, K. Kuepper, S. Dinklage, J. Wollschlaeger, L. Schneider, M. Steinhart, J. Hardege, D. Daum, *Energy Environ. Sci.,* 2015, **8***,* 2685-2697.
[42] H. Schäfer, S. M. Beladi-Mousavi, L. Walder, J. Wollschläger, O. Kuschel, S. Ichilmann, S. Sadaf, M. Steinhart, K. Küpper, L. Schneider, *ACS Catal.,* 2015, **5***,* 2671-2680.
[43] H. Schäfer, K. Küpper, J. Wollschläger, N. Kashaev , J. Hardege, L. Walder, S. M. Beladi-Mousavi, B. Hartmann-Azanza, M. Steinhart, S. Sadaf, F. Dorn, *ChemSusChem,* 2015, **21***,* 8, 3099-3110.
[44] Couto, S. R.; Sanroman, M. A.; Hofer, D.; Gubitz, G. M. Stainless steel sponge: a novel carrier for the immobilisation of the white-rot fungus Trametes hirsuta for decolourization of textile dyes. *Bioresour Technol* **2004**, *95*, 67-72.
[45] Liu, D.; Zheng, T.; Buisman, C.; Ter Heijne, A. Heat-Treated Stainless Steel Felt as a new Cathode Material in a Methane-Producing Bioelectrochemical System. *ACS Sustainable Chem. Eng.* **2017**, *5*, 11346-11353.
[46] Anantharaj, S.; Chatterjee, S.; Swaathini, K. C.; AmarnathT. S.; Subhashini, E.; Pattanayakm D. K.; Kundu, S. Stainless Steel Scrubber: A Cost Efficient Catalytic Electrode for Full Water Splitting in Alkaline Medium. *ACS Sustain. Chem. Eng.* **2017**, DOI:10.1021/acssuschemeng.7b03964
[47] H. Schäfer, K. Küpper, K. Müller-Buschbaum, D. Daum, M. Steinhart, J. Wollschläger, U. Krupp, M. Schmidt, W. Han, J. Stangl, *Nanoscale*, 2017, **9**, 17829-17838.
[48] A. Shinde, R. J. R. Jones, D. Guevarra, S. Mitrovic, N. Becera-Stasiewicz, J. A. Haber, J. Jin, J. M. Gregoire, *Electrocatalysis,* 2015, **6***,* 229-236.
[49] N. Suzuki, T. Horie, G. Kitahara, M. Murase, K. Shinozaki, Y. Morimoto, *Electrocatalysis,* 2016, **7***,* 115-120.
[50] MetalRavne, 2390 Ravne na Koroškem, Slovenija, EU. http://www.metalravne.com.
[51] S. Cherevko, S. Geiger, O. Kasian, N. Kulyk, J.-P. Grote, A. Savan, B. RatnaShresta, S. Merzlikin, B. Breitbach, A. Luwig, K. J. J. Mayrhofer, *Catal. Today,* 2016, **262***,* 170-180.
[52] M. Huynh, D. K. Bediako, D. G. Nocera, *J. Am. Chem. Soc.,* 2014, **136***,* 6002-6010.
[53] M. M. Najafpour, K. C. Leonard, F.-R. F. Fan, M. M. A. Tabrizi, A. J. Bard, C. K. King´ondu, S. L. Suib, B. Haghighi, S. I. Allakhverdiev, *Dalton Trans.,* 2013, **42***,* 5085-5091.
[54] L. C. Seitz, C. F. Dickens, K. Nishio, Y. Hikita, J. Montoya, A. Doyle, C. Kirk, A. Vojvodic, H. Y. Hwang, J. K. Norskov, T. F. Jaramillo, *Science,* 2016, **353***,* 6303, 1011-1014.
[55] P. P. Patel, M. K. Datta, O. I. Velikokhatnyi, R. Kuruba, K. Damodaran, P. Jampani, B. Gattu, P. M. Shanti, S. S. Damle, P. N. Kumta, *Scientific Rep.,* 2016, **6**:28367, DOI:10.1038/srep28367.
[56] R. Frydendal, E. A. Paoli, I. Chorkendorff, J. Rossmeisl, I. E. L. Stephens, *Adv. Energy Mater.,* 2015, **5***,* 1500991, 1-9.
[57] J. B. Gerken, J. G. McAlpin, J. Y. C. Chen, M. L. Rigsby, W. H. Casey, R. D. Britt, S. S. Stahl, *J. Am. Chem. Soc.,* 2011, **133***,* 14431-14442.
[58] R. L. Doyle, M. E. G. Lyons, *J. Solid State Electrochem.,* 2014, **18***,* 3271-3286.
[59] I. J. Godwin, R. L. Doyle, M. E. G. Lyons, *J. Electrochem. Soc.,* 2014, **161***,* F906-F917.
[60] Y. Lee, J. Suntivich, K. J. May, E. E. Perry, Y. Shao-Horn, *J. Phys. Chem Lett.,* 2012, **3***,* 399-404.
[61] M. E. G. Lyons, S. Floquet. *Phys. Chem. Chem. Phys.,* 2011, **13***,* 5314-5335.
[62] C. Iwakura, K. Hirao, H. Tamura, *Electrochim. Acta,* 1977, **22**, 335-340.
[63] S. Cherevko, A. R. Zeradjanin, A. A. Topalov, N. Kulyk, I. Katsounaros, K. J. J. Mayrhofer, *ChemCatChem,* 2014, **6***,* 2219-2223.
[64] M. Wohlfahrt-Mehrens, J. Heitbaum, *J. Electroanal. Chem.,* 1987, **237**, 251-260.
[65] O. Diaz-Morales, F. Calle-Vallejo, C. de Munck, M. T. M. Koper, *Chem. Sci.,* 2013, **4***,* 2334-2343.
[66] J. Wiltsau, O. Wolter, J. Heitbaum*, J. Electroanal. Chem*., 1985, **195**, 299-306.
[67] A. P. Grosvenor, B. A. Kobe, M. C. Biesinger, N. S. McIntrye, *Surf. Interface Anal.,* 2004, **36***,* 1564-1574.
[68] C. Klewe, M. Meinert, A. Boehnke, K. Kuepper, E. Arenholz, A. Gupta, J. M. Schmalhorst, T. Kuschel, G. Reiss, *J. Appl. Phys.,* 2014, **115***,* 123903, 1-7.

[69] M. C. Biesinger, B. P. Payne, A. P. Grosvenor, L. W. M. Lau, A. R. Gerson, R. St.C. Smart, *Appl. Surf. Sci.,* 2011*,* **257***,* 2717-2730.

[70] S. Uhlenbrock, C. Scharfschwerdt, M. Neumann, G. Illing, H. J. Freund, *J. Phys.: Condens. Matter.,* 1992, **4***,* 7973-7978.

[71] M. C. Biesinger, L. W. M. Lau, A. R. Gerson, R. St. C. Smart, *Phys. Chem. Chem. Phys.,* 2012, **14***,* 2434-2442.

[72] C. B. Lee, B. S. Kang, M. J. Lee, S. E. Ahn, G. Stefanovich, W. X. Xianyu, K. H. Kim, J. H. Hur, H. Yin, Y. Park, I. K. Yoo, J.-B. Park, B. H. Park, *Appl. Phys. Lett.,* 2007, **91***,* 082104, 1-3.

[73] D. K. Bediako, B. Lassalle-Kaiser, Y. Surendranath, J. Yano, V. K. Yachandra, D. G. Nocera, *J. Am. Chem. Soc.,* 2012, **134***,* 6801-6809.

[74] D. Friebel, M. W. Louie, M Bajdich, K. E. Sanwald, Y. Cai, A. M. Wise, M.-J. Cheng, D. Sokaras, T.-C. Wenig, R. Alonso-Mori, *J. Am. Chem. Soc.,* 2015, *137,* **3**, 1305-1313.

[75] E. Fabbri, A. Habereder, K. Walter, R. Kötz, T. J. Schmidt, *Catal. Sci. Technol.,* 2014*,* **4***,* 3800-3821.

[76] M. Huynh, D. K. Bediako, D. G. Nocera, *J. Am. Chem. Soc.,* 2014, **136***,* 6002-6010.

[77] A. T. Marshall, R. G. Haverkamp, *Electrochimica Acta,* 2010, **55***,* 1978-1984.

[78] W. Hu, S. Chen, Q. Xia. *Int. J. Hydrogen Energy,* 2014, **39**, 6967-6976.